\documentclass[aps,prd,twocolumn,groupedaddress,superscriptaddress]{revtex4-2}

\usepackage{lipsum}
\usepackage{graphicx}  
\usepackage{dcolumn}   
\usepackage{bm}        
\usepackage{amssymb}   
\usepackage{amsmath}
\usepackage{xcolor}
\usepackage{amsmath}
\usepackage{amsfonts}
\usepackage{hyperref}

\hypersetup{colorlinks=true, linkcolor=blue, citecolor=blue, urlcolor=blue}

\begin{document}


\title{\bf Search for Axion dark matter with the QUAX--LNF tunable haloscope}

\author{A.~Rettaroli}\email{alessio.rettaroli@lnf.infn.it} \affiliation{INFN, Laboratori Nazionali di Frascati, Frascati, Roma, Italy} 
\author{D.~Alesini} \affiliation{INFN, Laboratori Nazionali di Frascati, Frascati, Roma, Italy}
\author{D.~Babusci} \affiliation{INFN, Laboratori Nazionali di Frascati, Frascati, Roma, Italy}
\author{C.~Braggio} \affiliation{INFN, Sezione di Padova, Padova, Italy} \affiliation{Dipartimento di Fisica e Astronomia, Padova, Italy}
\author{G.~Carugno} \affiliation{INFN, Sezione di Padova, Padova, Italy}
\author{D.~D'Agostino} \affiliation{Dipartimento di Fisica E.R. Caianiello, Fisciano, Salerno, Italy} \affiliation{INFN, Sezione di Napoli, Napoli, Italy}
\author{A.~D'Elia} \affiliation{INFN, Laboratori Nazionali di Frascati, Frascati, Roma, Italy}
\author{D.~Di~Gioacchino} \affiliation{INFN, Laboratori Nazionali di Frascati, Frascati, Roma, Italy}
\author{R.~Di~Vora} \affiliation{INFN, Laboratori Nazionali di Legnaro, Legnaro, Padova, Italy}
\author{P.~Falferi} \affiliation{Istituto di Fotonica e Nanotecnologie, CNR Fondazione Bruno Kessler, I-38123 Povo, Trento, Italy} \affiliation{INFN, TIFPA, Povo, Trento, Italy}
\author{U.~Gambardella} \affiliation{Dipartimento di Fisica E.R. Caianiello, Fisciano, Salerno, Italy} \affiliation{INFN, Sezione di Napoli, Napoli, Italy}
\author{A. Gardikiotis} \affiliation{INFN, Sezione di Padova, Padova, Italy}
\author{C.~Gatti} \affiliation{INFN, Laboratori Nazionali di Frascati, Frascati, Roma, Italy}
\author{C.~Ligi} \affiliation{INFN, Laboratori Nazionali di Frascati, Frascati, Roma, Italy}
\author{A.~Lombardi} \affiliation{INFN, Laboratori Nazionali di Legnaro, Legnaro, Padova, Italy}
\author{G.~Maccarrone} \affiliation{INFN, Laboratori Nazionali di Frascati, Frascati, Roma, Italy}
\author{A.~Ortolan} \affiliation{INFN, Laboratori Nazionali di Legnaro, Legnaro, Padova, Italy}
\author{G.~Ruoso} \affiliation{INFN, Laboratori Nazionali di Legnaro, Legnaro, Padova, Italy}
\author{S.~Tocci} \email{simone.tocci@lnf.infn.it} \affiliation{INFN, Laboratori Nazionali di Frascati, Frascati, Roma, Italy}
\author{G.~Vidali} \affiliation{Dipartimento di Fisica, Universit{\`a} La Sapienza, Rome, Italy}\affiliation{INFN, Laboratori Nazionali di Frascati, Frascati, Roma, Italy}

\date{\today}

\begin{abstract}
We report the first experimental results obtained with the new haloscope of the QUAX experiment located  at Laboratori Nazionali di Frascati of INFN (LNF). The haloscope is composed of a OFHC Cu resonant cavity cooled down to about 30~mK and immersed in a magnetic field of 8~T. The cavity frequency was varied in a 6 MHz range between 8.831496 and 8.83803 GHz. This corresponds to a previously unprobed mass range between $36.52413$ and $36.5511~\mu\text{eV}$. We don't observe any excess in the power spectrum and set limits on the axion-photon coupling in this mass range down to $g_{a\gamma\gamma} \,\, \textless \,\, 0.882\times 10^{-13}$~GeV$^{-1}$  with the confidence level set to 90{\%}.

\end{abstract}


\maketitle


\section{\label{sec:intro}Introduction}
In recent years, we witnessed an increasing growth in the research of light Dark Matter (DM) candidates, addressing in particular axions and axion-like particles (ALPs). If axions are found to exist, they would untie the long-standing DM problem~\cite{weinberg1978new,wilczek1978problem}, after being originally postulated as a solution to the strong CP problem~\cite{peccei1977cp,peccei1977cp_2}. The nature of a pseudoscalar, electrically neutral and feebly interacting particle makes the axion a strong DM candidate~\cite{preskill1983cosmology,abbott1983,dine1983}, and its cosmological evolution and astrophysical constraints indicate a favorable mass range between $1~\mu\text{eV} < m_a < 10~\text{meV}$~\cite{irastorza2018new}.

The research efforts are now spread over many different detection approaches, but the paradigm has become the haloscope design proposed by Sikivie~\cite{sikivie1983experimental,PhysRevD.32.2988}, which probes axions from the DM halo of the Galaxy. Currently operating haloscopes are ADMX~\cite{du2018search,boutan2018piezoelectrically,braine2020extended,bartram2021search}, HAYSTAC~\cite{zhong2018results,backes2021quantum,jewell2023}, ORGAN~\cite{mcallister2017organ,quiskamp2022}, CAPP-8T~\cite{lee2020axion,choi2021capp}, CAPP-9T~\cite{jeong2020search}, CAPP-PACE~\cite{kwon2021first}, CAPP-18T~\cite{lee2022searching}, CAST-CAPP~\cite{castcapp2022}, CAPP-12TB~\cite{yi2023capp}, GrAHal~\cite{grenet2021grenoble}, RADES~\cite{melcon2018axion,melcon2020scalable,alvarez2021first}, TASEH~\cite{chang2022first} and QUAX~\cite{barbieri2017searching,crescini2018operation,alesini2019galactic,crescini2020axion,alesini2021search,alesini2022search,DiVora2023search}, whereas among the proposed experiments are FLASH~\cite{flashCDR}, BabyIAXO/RADES~\cite{babyiaxo-rades}, ABRACADABRA~\cite{ouellet2019}, DM-Radio~\cite{brouwer2022m3,brouwer2022gut}, CADEx~\cite{aja2022}, MADMAX~\cite{caldwell2017dielectric}, ALPHA~\cite{lawson2019tunable}, WISPLC~\cite{wisplc}, DALI~\cite{dali}, BRASS~\cite{brass}, BREAD~\cite{bread} and SUPAX~\cite{supax}.

The axion observation technique is based upon its inverse Primakoff conversion into one photon, stimulated by a static magnetic field. The essential elements required to run a haloscope are a superconducting magnet to generate a strong magnetic field, a microwave resonant cavity where the electromagnetic field excitation builds up, an ultra-low noise receiver, a tuning mechanism to scan over the axion mass range and a cryogenic system to grant operation at low temperature. The two figures of merit in the axion search are the power of the produced photon~\cite{BrubakerPRL}
\begin{equation}\begin{split}
	\label{eq:power}
	P_{a\gamma}=\left( \frac{g_{a\gamma\gamma}^2}{m_a^2}\, \hbar^3 c^3\rho_a \right)
	\left( \frac{\beta}{1+\beta} \omega_c \frac{1}{\mu_0} B_0^2 V C_{010} Q_L \right)& \\
    \times \left(\frac{1}{1+\left(2Q_L \, \Delta \omega_c  / \omega_c \right)^2}\right),
	\end{split}
    \end{equation}
and the scan rate~\cite{Kim2020scanrate}
\begin{equation}\begin{split}
	\label{eq:scanrate}
	\frac{\text{d}f}{\text{d}t}= g_{a\gamma\gamma}^4 \, \frac{\rho_a^2}{m_a^2} \, \frac{1}{\text{SNR}^2} \left( \frac{B_0^2VC_{010}}{k_B T_{\text{sys}}} \right)^2 & \\%
    \times \frac{\beta^2}{\left( 1+\beta \right)^2} \, Q_a \, \text{min}\left(Q_L,Q_a\right).
    \end{split}
    \end{equation}
We assume a local DM density $\rho_a = 0.45~\text{GeV}/\text{cm}^3$~\cite{10.1093/ptep/ptaa104}, $m_a$ is the axion mass and $g_{a\gamma\gamma}$ is its coupling constant to photons. $B_0$ is the applied magnetic field; $\omega_c = 2\pi\nu_c$, $V$, $Q_L$, $\beta$ are respectively the resonance angular frequency of the cavity, the volume, the loaded quality factor and the antenna coupling to the cavity. The relation $Q_L = Q_0/\left(1+\beta\right)$ holds, with $Q_0$ the intrinsic quality factor. $C_{010}$ is a mode dependent geometrical factor, about 0.69 for the TM010 resonant mode of a cylindrical cavity, and $\Delta \omega_c = \omega_c - \omega_a$ is the detuning between the cavity and the axion angular frequency defined as $\omega_a=m_a c^2/\hbar$. The quality factor $Q_a \approx 10^6$~\cite{turner1990periodic} is related to the energy spread in the cold dark matter halo. In Eq.~\eqref{eq:scanrate}, the signal-to-noise ratio, SNR, is defined as the ratio between the signal power and the uncertainty of the noise power, $\text{SNR} \equiv P_{a \gamma}/\delta P_{\text{noise}}$, while $T_{\text{sys}}$ is the system noise temperature, whose reduction is of fundamental importance in speeding up the scan rate of the experiment.

In this paper, we report on the first operation of the new QUAX haloscope located at the National Laboratories of Frascati (LNF), which doubles the search potential of the QUAX experiment along with the LNL haloscope~\cite{DiVora2023search}. The experiment is conducted using a resonant cavity equipped with a tuning rod mechanism allowing to exclude the existence of dark matter axions with coupling $g_{a\gamma\gamma}$ down to $0.882\times 10^{-13}~\text{GeV}^{-1}$ in the mass window $(36.5241-36.5510)~\mu\text{eV}$.

\section{\label{sec:setup}Experimental Setup}

\subsubsection{\label{sec:haloscope}Haloscope description}
\begin{figure}
  \centering
      \includegraphics[width=0.47\textwidth]{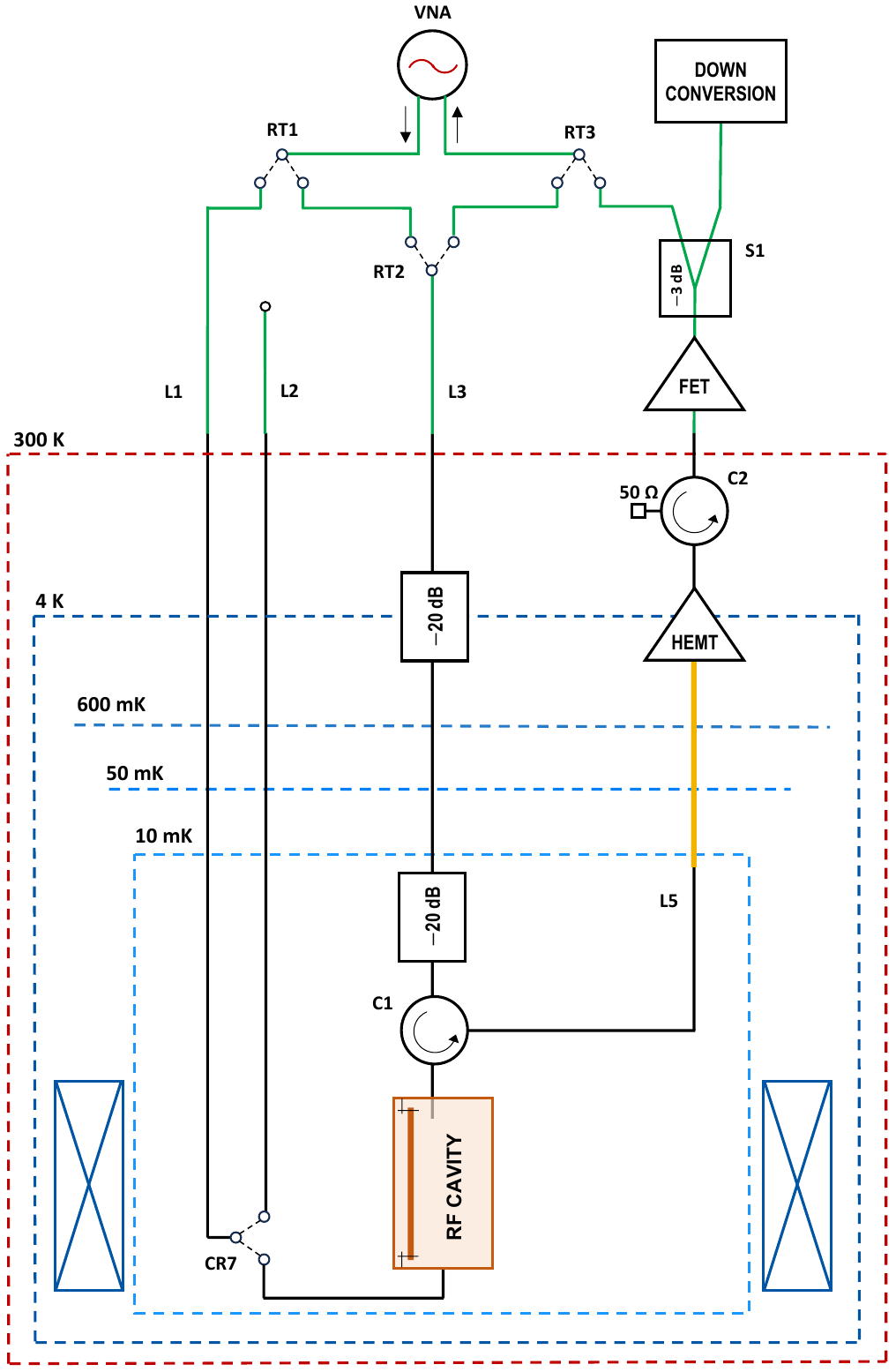}
\caption{\small Experimental setup sketch. The successive temperature stages of the dilution refrigerator are shown with their base temperature. L1 and L3 are input RF lines, L5 is the output amplified line and L2 is an auxiliary line used for further checks. Note that L1, L2 and L3 have intrinsic attenuations of about 15~dB each. The room temperature switches (RT1, RT2, RT3) and the cryogenic switch (CR7) allow all the combinations required for measurements and calibrations. Attenuators are indicated with their attenuation values, while C1 and C2 are circulators. A superconductive RF cable in L5 is indicated in yellow. The magnet is shown as two crossed squares and is thermalized in the 4~K vessel. The HEMT and FET are, respectively, the cryogenic and room-temperature amplifiers, whereas the power splitter is indicated as S1.}
\label{fig:RFdiagram}
\end{figure}

The haloscope, schematized in Fig.~\ref{fig:RFdiagram}, consists of a cylindrical OFHC copper resonant cavity, with inner radius $r=13.51$~mm and height $h=246$~mm, for a total volume $V=0.141$~l. The body is divided into two semi-cylinders, including the endcaps, sealed together with screws. Two coaxial cables are coupled to the cavity via dipole antennas. One antenna is fixed and weakly coupled, with coupling estimated from simulations $1.4\times 10^{-3}$ and verified to be less than $7\times 10^{-3}$ from calibration data, such that Eqs.~\eqref{eq:power} and~\eqref{eq:scanrate} are still valid. The other is obtained by stripping the end of a coax cable, leaving only the central conductor for a 2 mm length (Bottom-Left panel of Fig.~\ref{fig:rod}). The coax cable is connected to a linear stepper motor whose movement allows the tunability of the  coupling $\beta$.

The magnetic field is provided by a NbTi superconducting solenoid magnet with a cold bore diameter of 100~mm and 320~mm height. On top of the magnet, a second coil assures the reduction of the stray field above the cavity below few hundred Gauss. The magnet, initially operated at the nominal field of 9~T, was set to a lower safety value following a quench during a current ramp. The experiment here described was then conducted at 8~T, with a field inside the cavity volume of r.m.s. $\sqrt{\langle B_0^2\rangle}=6.73$~T.

The power collected through the tunable antenna is amplified along the output line L5 (see Fig.~\ref{fig:RFdiagram}) by the first amplification stage consisting in a cryogenic high electron mobility transistor (HEMT) amplifier. The circulator C1 allows the routing of the RF signals  and provides the isolation of the cavity from the HEMT. The HEMT output signal reaches a room temperature field-effect (FET) amplifier through a second circulator, C2, and is then split and transmitted to the downconversion electronics or to a Vector Network Analyzer (VNA) by means of the switch S1.

The setup is hosted in a Leiden Cryogenics CF-CS110-1000 dilution refrigerator equipped with two Sumitomo pulse tubes with cooling power of 1.5 W at 4 K each. The refrigerator is segmented into different temperature stages. The magnet and the HEMT are thermalized at the 4~K stage, while the cavity is connected to the last temperature stage which attained 20~mK at equilibrium. The cavity temperature, monitored during the data taking, reached about 30~mK.

\subsubsection{\label{sec:tuning}Tuning mechanism}
\begin{figure}
  \centering
      \includegraphics[width=0.485\columnwidth]{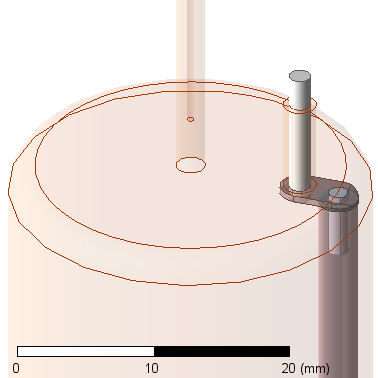}
      \includegraphics[width=0.485\columnwidth]{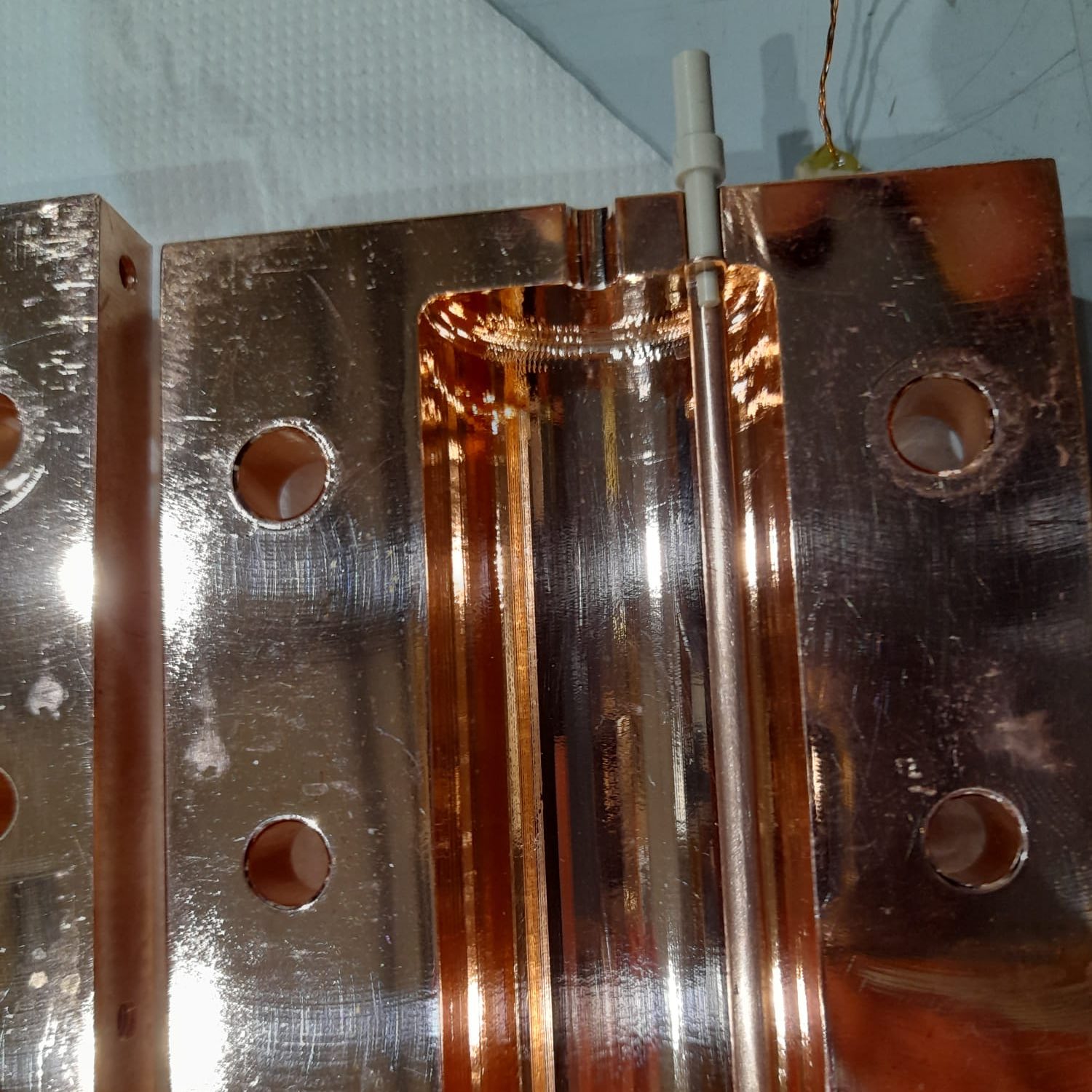}\\
    \includegraphics[width=0.485\columnwidth]{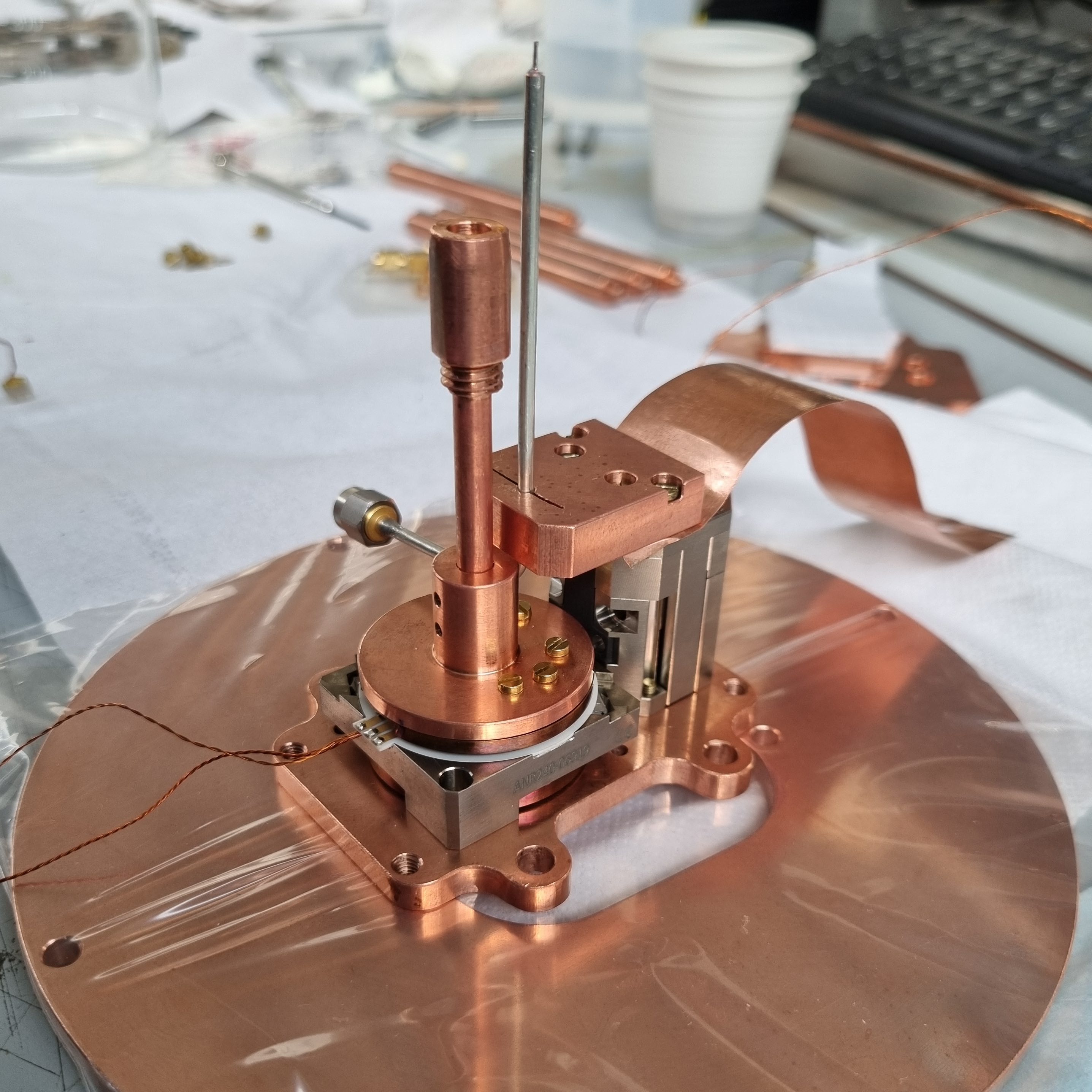}
    \includegraphics[width=0.485\columnwidth]{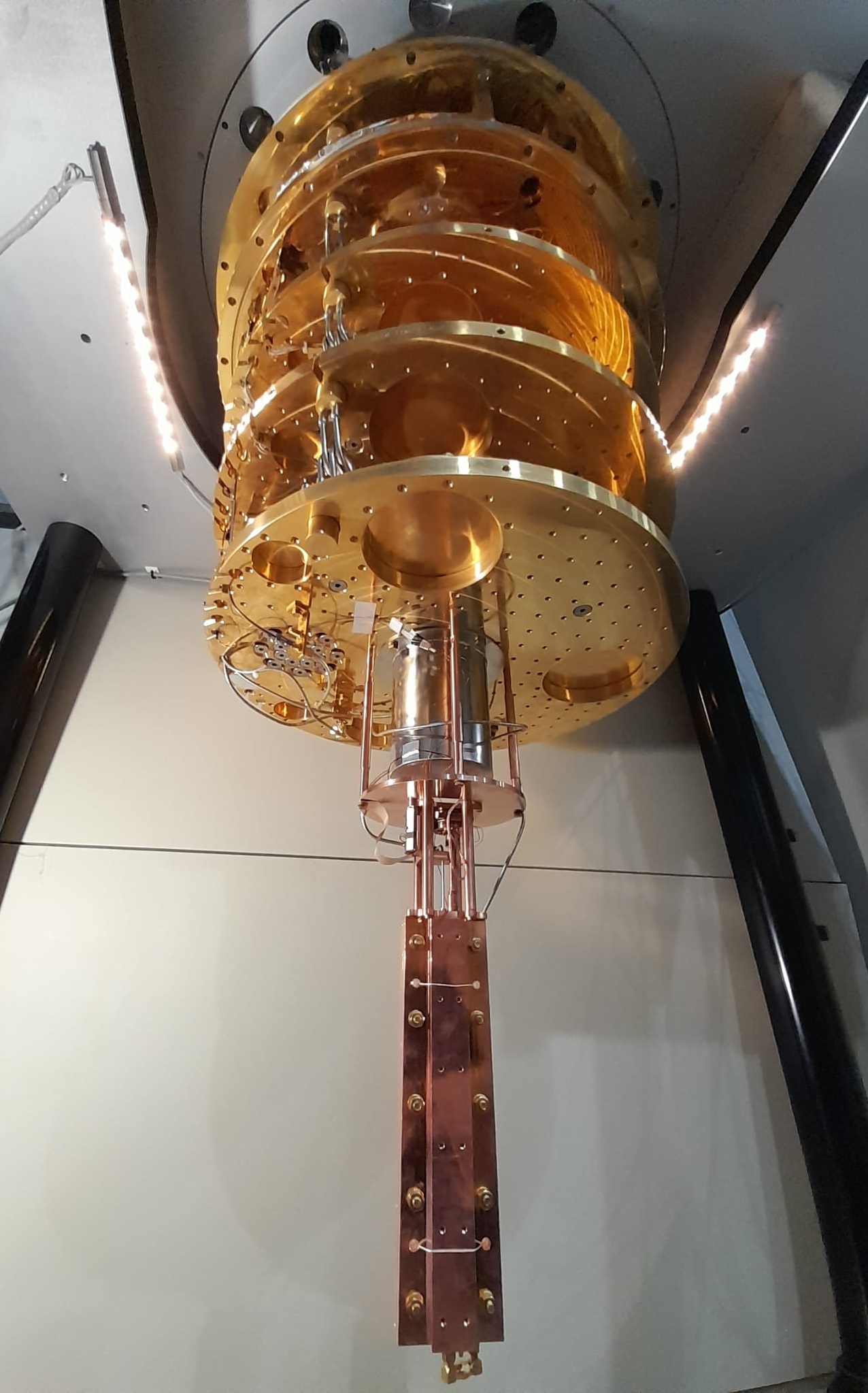}
\caption{\small Experimental setup closeups. \emph{Upper left)} Detail of the rod tuning mechanism design as seen in the simulator. \emph{Upper right)} Picture of one end of the rod with the PEEK support, placed in one of the two cavity halves. \emph{Bottom left)} Stepper motors detail. On the left is the rotative motor together with the mandrel to cling the rod support, on the right is the linear motor holding the movable antenna. \emph{Bottom right)} View of the assembled haloscope, with the resonant cavity attached to the 20~mK stage with copper bars.}
\label{fig:rod}
\end{figure}

The frequency tuning is obtained by moving a copper rod with radius 1.5~mm and length 244~mm inside the cavity volume (Fig.~\ref{fig:rod}). The effective cavity volume is then reduced to $V=0.139$~l. The rod is supported by PEEK nails, which are centered off-axis with respect to the rod, as shown in Fig.~\ref{fig:rod}. At one end, the PEEK is grabbed by a copper mandrel, which is rotated by the stepper motor. Thanks to this movement, the rod accomplishes an arc of circumference approaching the center of the cavity. The electromagnetic behavior of this system was simulated with the ANSYS HFSS suite~\cite{hfss}. The resonant mode of interest, TM010, has a starting frequency of 8.817~GHz when the (ideal) rod is at rest in contact with the cavity wall. Moving the rod towards the center, the mode is squeezed and simulations indicate that the frequency is tuned up to 9.106~GHz with a rotation of 80 degrees, while keeping the geometric factor $C_{010}$ close to its ideal value and with a reduction of the quality factor of about 10\%.

The rod orientation is changed by a rotative stepper piezoelectric motor from Attocube Systems model ANR240, while the second linear motor, as anticipated, moves the tunable antenna (model ANPz111). Both motors were operated at cryogenic temperatures, between 20 and 30 mK, without complications, and with a heating up of the environment of only a few~mK when operated in single-step mode.

\begin{figure}
    \centering
      \includegraphics[width=0.95\columnwidth]{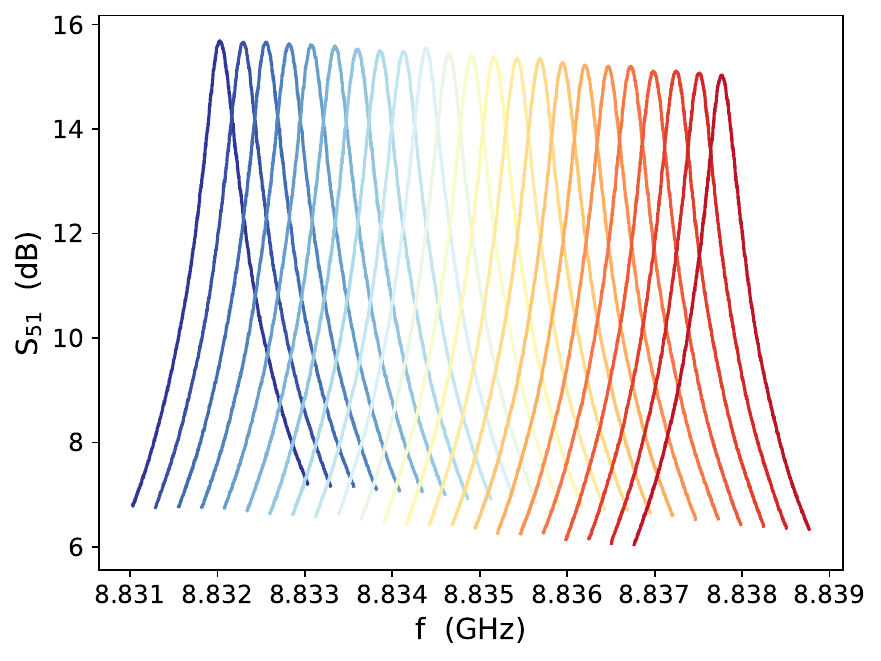}
\caption{\small Transmission spectra ($S_{51}$) of the haloscope cavity for different resonant frequencies. We tuned the resonance of the cavity in the 6~MHz frequency window during the axion search.} \label{fig:scan} 
\end{figure}

Although the intrinsic quality factor of the cavity without the tuner is measured to be $Q_0\simeq 10^5$ at cryogenic temperatures, the PEEK supporting the rod and non-ideality of the rod itself caused field losses, reducing the quality factor to about 50000 (Sec.~\ref{sec:datataking}).
The coupling, supposed to be equal to the optimal value for scan $\beta=2$, resulted to be $\beta=0.5$ after a more accurate analysis of calibration data.
The values of $Q_L$ and $\beta$ measured before the data taking at different frequencies are listed in Tab.~\ref{tab:dataset}, while the transmission spectra measured with the VNA are shown in Fig.~\ref{fig:scan}.

\subsubsection{\label{sec:datataking}Data taking}

After the cool-down, the following procedure is used for calibrating and taking data at each frequency step, starting from the first measured value of 8.8317690~GHz.
\begin{itemize}
    \item The cavity frequency is set moving the ANR240 motor step by step while monitoring the $S_{51}$ from the VNA. Each time the frequency is increased by one cavity bandwidth, about 260~kHz, with respect to the previous data taking.
    \item The waveforms of the scattering parameters $S_{51}$, $S_{53}$ and $S_{13}$, named according to the line numbering in Fig.~\ref{fig:RFdiagram}, are collected with the VNA, to perform the calibration.
    \item The raw data acquisition is started, lasting about 3600 s.
\end{itemize}

The data acquisition consists in recording the power coming from the output line L5.
During data acquisition, the switch CR7 is closed on line L2 to avoid noise leakage through the under coupled port, the switch RT1 is closed on the line L1 and the switch RT2 is closed on the switch RT3 which, in turn, is closed on a 50 ohm termination (not shown in Fig.~\ref{fig:RFdiagram}). Finally, the splitter S1 allows to redirect the signal to the downconversion and acquisition electronics. Here, with an I--Q mixer we convert the frequency to the baseband, and the I and Q quadratures are amplified by low-noise voltage amplifiers ($\times 10^3$ factor, 10~MHz bandwidth) before being digitized by a 16-bit ADC board, which has a 2~MHz bandwidth and sampling of 2~MS/s. In each sub-run, the I and Q signals are acquired for 4 seconds and saved to file. The total amount of files in each sub-run is 941, resulting in an integration time of $\Delta t = 3764$~s.

\begin{widetext}
    \begin{figure*}[t]
    \centering
      \includegraphics[width=0.98\textwidth]{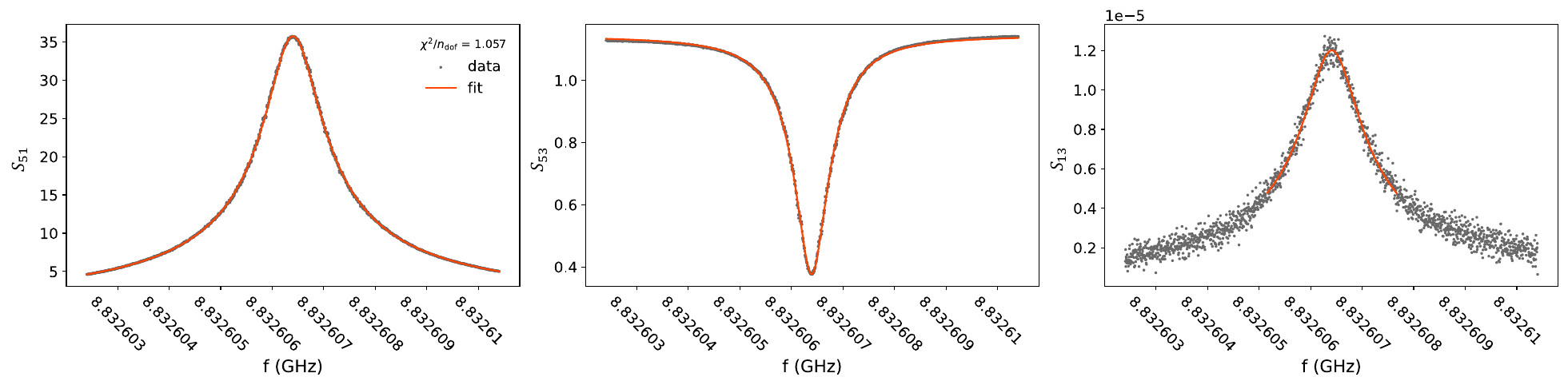} \caption{\small Fit example of the scattering parameters $S_{51}$ (forward transmission), $S_{53}$ (reflection) and $S_{13}$ (backward transmission) to extract the cavity parameters and gain in the calibration.}
        
\label{fig:calib}
\end{figure*}
\end{widetext}

The calibration procedure (refer to Appendix~\ref{sec:calib} for more details) is done by a simultaneous fit of the $S_{51}$, $S_{53}$ and $S_{13}$ spectra with their analytical expressions, allowing the extraction of the cavity parameters $\nu_c$, $Q_0$, $\beta$ and of the attenuation and gain of the input and output lines (see Fig.~\ref{fig:calib}). In particular, we are interested in the gain of Line~5, which is calculated solving a system of three coupled equations (Eq.~\eqref{eq:sistemaEq}).
The measured gain through all the sub-runs is
\begin{equation}
 G_{L5} = \left( 70.62 \pm 0.28_{\scriptscriptstyle \text{syst.}} \pm 0.13_{\scriptscriptstyle \text{stat.}} \right)~\text{dB}, 
 \label{eq:gain5}
\end{equation}
where the systematic uncertainty is mainly due to the gain flatness of the reconstracted $G_{L5}$, and the statistic uncertainty derives from the data scattering of the $S_{13}$ trace. This gain is referred at the input of the splitter, having subtracted the cable contributions from the splitter to the VNA and the $-3$~dB of the splitter itself. Then, the gain spectrum from the splitter output to the ADC is measured, taking into account the effect of the ADC internal filters.

\section{\label{sec:results}Data Analysis and Results}
The cavity resonance frequency $\nu_c$ is tuned in a 6~MHz range between 8.831769--8.8377664 GHz in 24 steps. The parameters of each of the 24 datasets are reported in (Tab.~\ref{tab:dataset}). For each dataset we calculate the power spectrum by combining the quadratures as $I - iQ$, computing the FFT and taking the squared module. The spectra are centered at the LO frequency, which is always $\nu_{\text{LO}} = \nu_c - 500~\text{kHz}$, and are corrected for the Line~5 gain $G_5$ and for the spectrum shape of the downconversion electronics.

\begin{table}[tb]
  \begin{center}
    \caption{Cavity resonance frequency, quality factor and cavity-antenna coupling for each dataset of the scan.}
    \label{tab:dataset}
  \vspace*{0.5cm}
    \begin{tabular}{c|c|c}
			\hline\hline
            $\nu_c$ [GHz]& $Q_L$ &  $\beta$   \\\hline
                8.83176900 & 32345 & 0.5206  \\
			8.83203080 & 32228 & 0.519\\
			8.83229550 & 32273 & 0.5082\\
			8.83255580 & 32332 & 0.5141\\
			8.83282190 & 32387 & 0.5097 \\
			8.83307310 & 32401 & 0.5078\\
			8.83334500 & 32300 & 0.5097 \\
                8.83360070 & 32503 & 0.5058\\
			8.83386200 & 32540 & 0.5075 \\
			8.83412790 & 32752 & 0.5014\\
			8.83438580 & 32573 & 0.5026 \\
                8.83464620 & 32904 & 0.5005 \\
                8.83490660 & 32957 & 0.4984 \\
                8.83516350 & 32863 & 0.4951 \\
                8.83542850 & 32872 & 0.4947 \\
                8.83568970 & 33326 & 0.4881 \\
                8.83594630 & 33051 & 0.489 \\
                8.83620570 & 33056 & 0.4894 \\
                8.83646975 & 33104 & 0.4857 \\
                8.83672330 & 33584 & 0.4823 \\
                8.83698660 & 33529 & 0.4803 \\
                8.83724500 & 33659 & 0.4823 \\
                8.83750860 & 33639 & 0.4793 \\
                8.83776640 & 33450 & 0.4793\\
                
			\hline
			\hline
    \end{tabular}
  \end{center}
\end{table}

For each frequency step, we estimate $T_{\rm sys}$ from the output power, properly converted into units of Kelvin, at a reference frequency $\nu_{\rm LO}+ 100~\text{kHz}$, where 1/f noise is negligible and still far enough from the cavity resonance, obtaining an average value of 4.7~K. Across the dataset we observe that the power spectra line shapes show features typical of Fano interference, as recently reported in~\cite{rieger2023fano} where similar interference phenomena  in microwaves measurements were analyzed. We interpret the bump structure as due to a imperfect isolation and thermalization of the circulator, attenuators and the rod inside the cavity, leading to effective temperatures greater than expected from thermalization; in particular we estimate $T_{\textup{circ}}\approx 600$~mK and $T_{\textup{cav}}\approx 400$~mK. This effect does not affect the extraction of the power generated by axion conversion since it adds incoherently.

\begin{figure}
  \centering
      \includegraphics[width=0.47\textwidth]{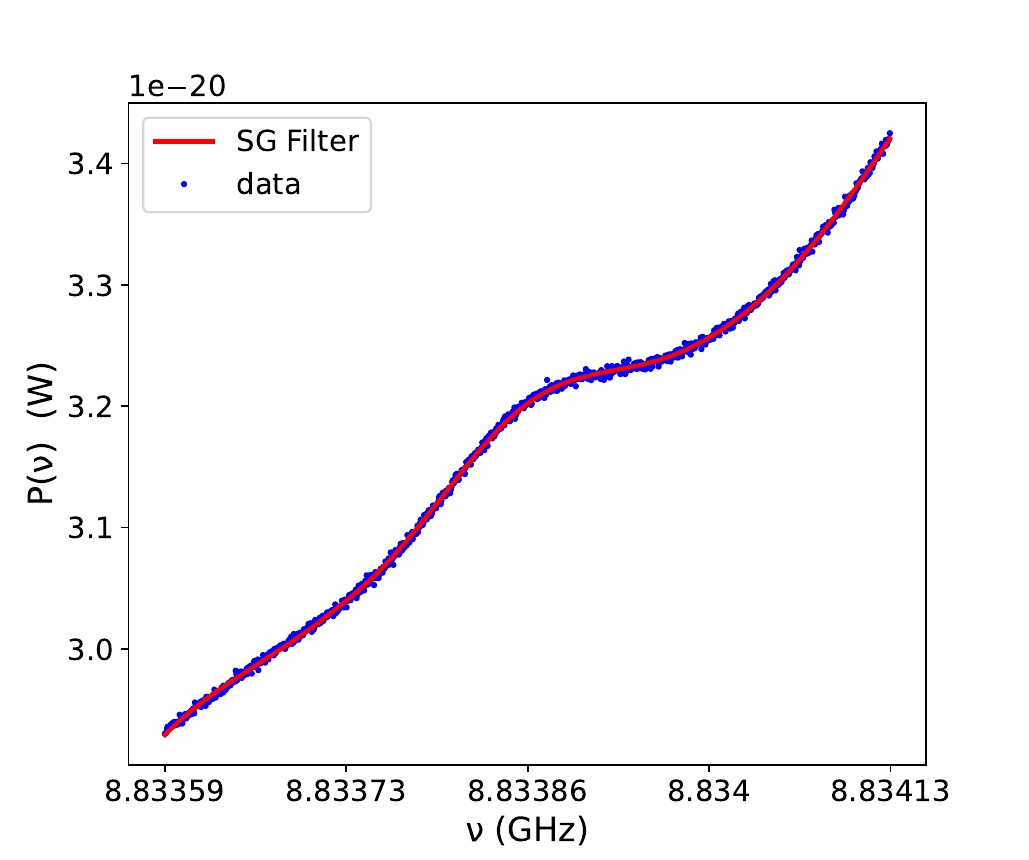}
\caption{\small FFT cavity power spectrum (blue dots) and SG filter (red line). $\nu_c=8.833862$ GHz, $Q_L=32540$.}
\label{fig:SG+data}
\end{figure}

\begin{figure}[t]
  \centering
     \includegraphics[width=0.47\textwidth]{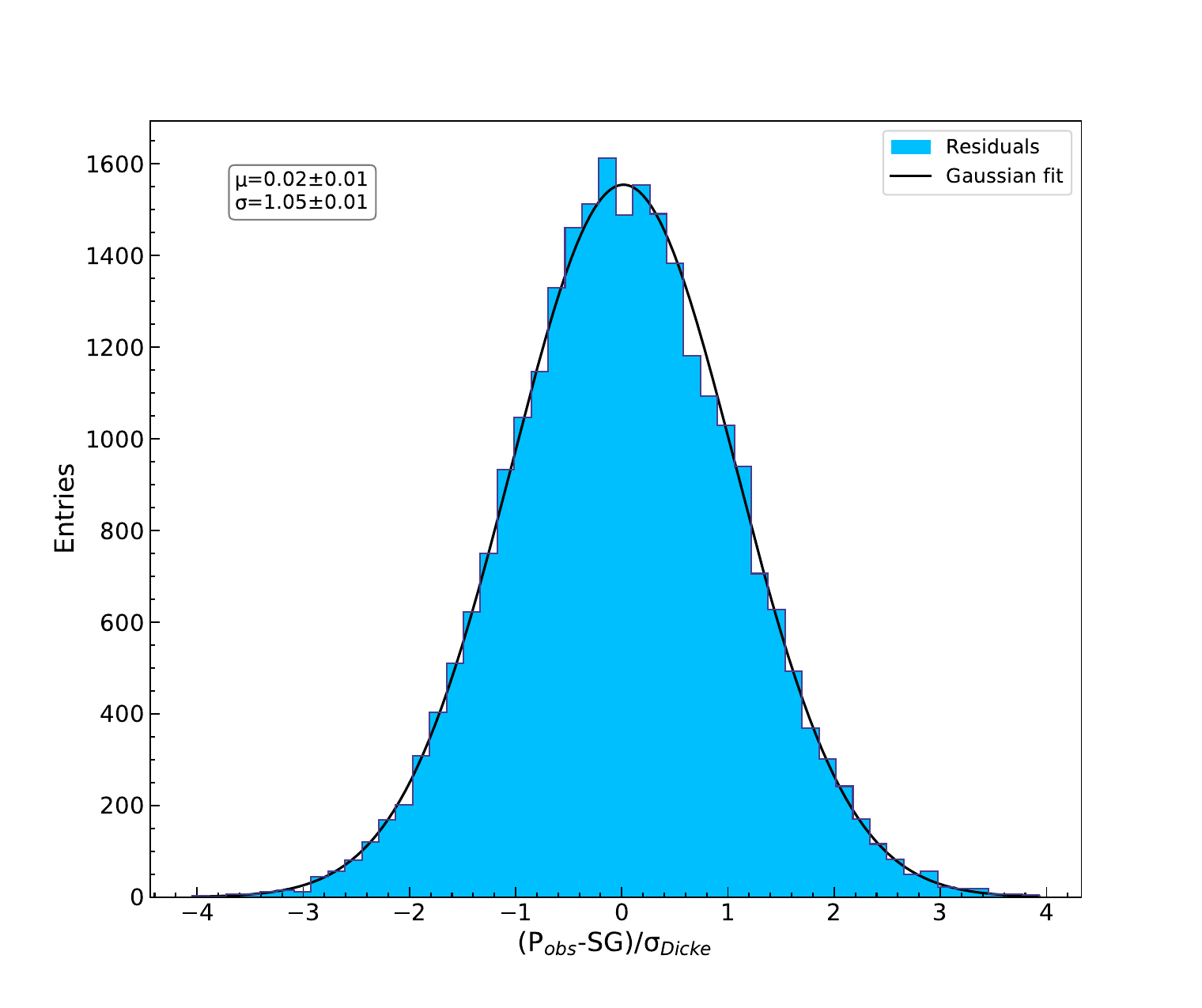}
\caption{\small Distribution of the cumulative residuals normalized to 
$\sigma_{\scriptscriptstyle \textup{Dicke}}$.}
\label{fig:cumulative_residual}
\end{figure}

In presence of axion conversions a power surplus is expected in the cavity power spectrum as given by Eq.~\eqref{eq:power}. We calculate the power spectrum residuals by subtracting it to a polynomial obtained with a Savitzky-Golay (SG) filter~\cite{savitzky1964smoothing} from the spectrum itself. We adopt a SG filter of the fourth order with a dynamic interval of 250.5 kHz (501 bins). We apply the SG filter in a window [$\nu_c-\Gamma, \nu_c+\Gamma$] of about 530 kHz, for each dataset. $\Gamma$ is the cavity line-width and is calculated as $\nu_c/Q_L$.  An example of polynomial obtained from a SG filter from one of the datasets is shown in Fig.~\ref{fig:SG+data}. The slope in the power spectrum is due to aliasing of the thermal noise at the edge of the Nyquist window, while the bump structure is due to the imperfect thermalization and isolation of the rf components connected to the resonant cavity, as pointed out earlier.
We normalize the residuals to the expected noise power $\sigma_{\scriptscriptstyle \textup{Dicke}}$ for each dataset, where $\sigma_{\scriptscriptstyle \textup{Dicke}}$ is calculated using the Dicke radiometer equation ~\cite{dicke1946measurement}

\begin{equation}
  \sigma_{\scriptscriptstyle \textup{Dicke}} = k_B T_{\rm sys} \sqrt{\Delta \nu/\Delta t}\, ,
\end{equation}

where $T_{\rm sys}$, is the system noise temperature, $\Delta \nu$ is the bin width (500 Hz) and $\Delta t$ is the integration time (3764 s).
The distribution of the normalized residuals obtained combining all the datasets, is shown in Fig.~\ref{fig:cumulative_residual}. The data follow a Gaussian distribution, with a standard deviation compatible with 1.

For each axion mass, we apply the Least-Squares method to estimate the best value $\hat{g}_{a\gamma\gamma}$ by minimizing    
\begin{equation}
 \chi^2 = \sum_{\alpha=1}^{N_{\text{scan}}} \sum_{i=1}^{N_{\text{bin}}}\,
\left[\frac{R_{\alpha,i} - S_{\alpha,i} (m_a, g_{a\gamma\gamma}^2)}{\sigma^{(\alpha)}_{\text{Dicke}}}\right]^2\,,
\label{eq:chisquare}
\end{equation}
where the $\alpha$ index runs over the number of datasets ($N_{\text{scan}}$), the index $i$ runs over the frequency bins of each power spectrum. $S_{\alpha,i}$ and $R_{\alpha,i}$ are the expected power signal and the residuals for frequency bin $i$ and the dataset $\alpha$. We calculate $S_{\alpha,i}$ as the integral in the frequency domain of Eq.~\eqref{eq:power} multiplied by the spectrum of the full standard halo model distribution \cite{turner1990periodic}.

Expressing $S_{\alpha,i} (m_{a},g_{a\gamma\gamma}^2) = g_{a\gamma\gamma}^2\,T_{\alpha,i} (m_{a})$, we analytically minimize Eq.~\eqref{eq:chisquare} by solving $\partial \chi^2 / \partial g^2_{a\gamma\gamma} = 0$, and 
calculate the uncertainty according to the formula $(\xi = g^2_{a\gamma\gamma})$:
\begin{equation}
\frac1{\sigma^2_{\hat{\xi}}} = \frac12\,\frac{\partial^2 \chi^2}{\partial^2 \xi}\,.
\end{equation}

Minimizing Eq.~\eqref{eq:chisquare} we find: ($\sum\sum \equiv \sum_{\alpha=1}^{N_{\text{scan}}} \sum_{i=1}^{N_{\text{bin}}}$)
\begin{equation}
  \overline{g^2} =  \sigma^2 (\overline{g^2})\,\left[\sum\sum\,\frac{R_{\alpha,i}\,T_{\alpha,i} (m_a)}{(\sigma^{(\alpha)}_{\text{Dicke}})^2}\right]\,
\label{eq:G2}
\end{equation}
where $\overline{g^2}$ is the average squared coupling-constant that incorporates the contributions of all the frequency bins from all the datasets, and
\begin{equation}
\sigma^2 (\overline{g^2}) = \sum\sum \, \left[\frac{T_{\alpha,i} (m_a)}{\sigma^{(\alpha)}_{\text{Dicke}}}\right]^2
\label{eq:sigmaG2}
\end{equation}
is its variance. 
We repeat this procedure for different values of $m_a$ in the range $36.5241-36.5510$ $\mu$eV. 
The detection of a power excess larger than 5$\sigma$ above the noise is required for a candidate discovery. Since no candidates are found (Fig. \ref{fig:cumulative_residual}), we determine the exclusion limits for $g_{a\gamma\gamma}$ in this mass range as follows.

To correctly estimate these limits, we account for the efficiency of the filtering procedure used for the extraction of the axion signal. We follow the procedure reported in \cite{alesini2022search}. We run a Monte Carlo simulation numerically injecting a fake axion signal, with a known $g_{\text{injected}}^2$, into simulated power-spectra with different $\nu_c$. We use Eq.~\eqref{eq:G2} to estimate the $g_{\text{injected}}^2$, for each injected signal (i.e. obtaining $g_{\text{calculated}}^2$). Hence, we determine the efficiency from the relation between $g_{\text{calculated}}^2$  and $g_{\text{injected}}^2$.
To simulate the cavity power-spectra we add random Gaussian noise (with mean=0, sigma=$\sigma_{\scriptscriptstyle \text{Dicke}}$, according to a Gaussian PDF) to the SG filters.  Applying this procedure we estimate a detection efficiency of 0.845 on $g_{\text{calculated}}^2$ for the SG filter. After correcting for the estimated efficiency, we calculate the single-sided upper limit on the axion-photon coupling with a 90\% confidence level as in~\cite{alesini2021search}. We use a power constrained procedure for the $\overline{g^2}$ that under fluctuates below $-\sigma$ ~\cite{cowan2011power}. 
\begin{figure}[t]
  \centering
      \includegraphics[width=0.48\textwidth]{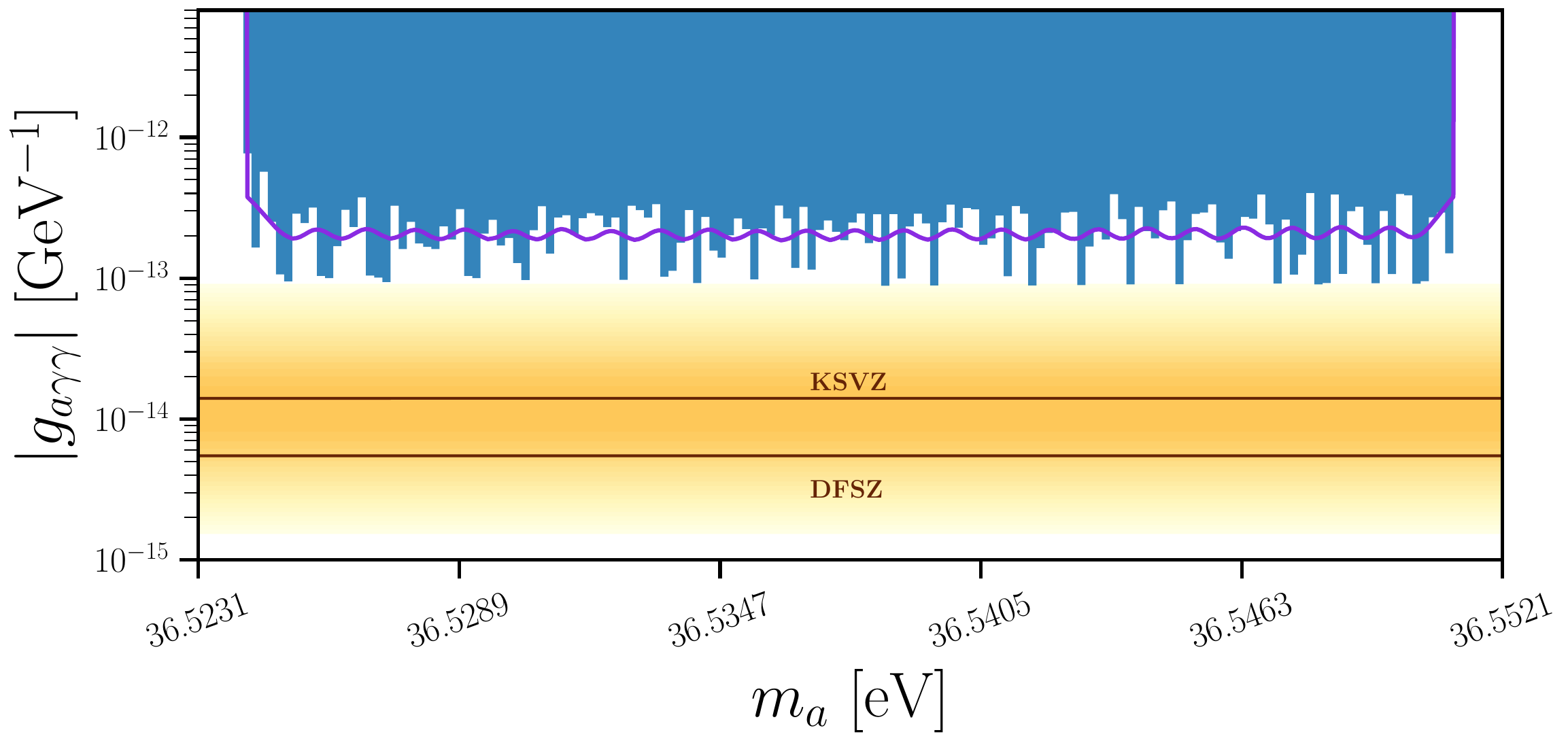}
      \includegraphics[width=0.48\textwidth]{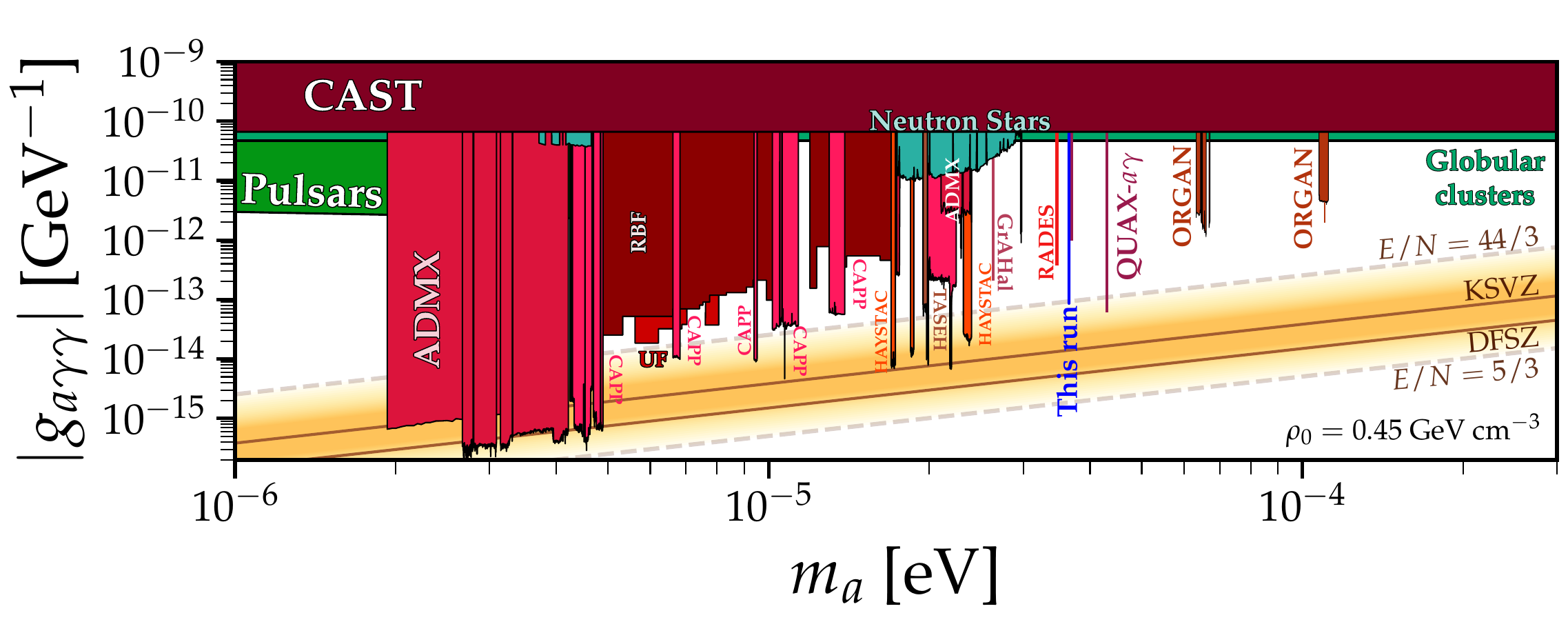}
\caption{\small \emph{Top)} The 90\% single-sided C.L. upper limit for $g_{a \gamma \gamma}$ as a function of the axion mass. The violet solid curve represents the expected limit in the case of no signal. The yellow region indicates the QCD axion model band. \emph{Bottom)} Broad view of axion exclusion limits set by haloscopes. The blue vertical line indicates the search presented in this paper. Images realized with~\cite{AxionLimits}. }
\label{fig:gCL}
\end{figure}
In Fig.~\ref{fig:gCL} we show the upper-limit  $g_{a\gamma\gamma}^{\scriptscriptstyle \textup{CL}}$ in an axion mass window of  27.02 neV  centered around  $36.53764~\mu$eV. The maximum sensitivity obtained with a 90\% CL is $g_{a\gamma\gamma}^{\scriptscriptstyle \textup{CL}} < 0.882\times10^{-13}$~GeV$^{-1}$.
This value is about 6 times larger with respect to the benchmark QCD axion level of the KSVZ theory~\cite{ksvz1979,ksvz1980}.

\section{\label{sec:conclusions}Conclusions}
In this paper we described the first search for axion dark matter with the tunable QUAX--LNF haloscope. The operation of a microwave cavity resonating at 8.8 GHz and equipped with a movable tuning rod in a magnetic field of 8~T, placed in a stable ultra-cryogenic environment, demonstrated to be successful. Taking data with this setup for a total amount of about 25~hours through 6~MHz allowed us to exclude the existence of axion dark matter in a mass range between $36.52413$ and $36.5511~\mu\text{eV}$ with axion-photon coupling $g_{a\gamma\gamma}$  down to $0.882\times 10^{-13}$~GeV$^{-1}$ with a C.L. of 90{\%}.

The apparatus has still much room for improvement. In the upcoming future, the sensitivity can be boosted employing a superconducting resonant cavity, enhancing the intrinsic quality factor. Adding a Josephson Parametric Amplifier as a preamplification stage, which importance have been widely demonstrated in this field, will break the noise temperature down by one order of magnitude. Moreover, the rod design can be optimized: \emph{i.} reducing the amount of dielectric exposed to the electromagnetic mode of the cavity, \emph{ii.} placing the lower end of the bar directly in contact with the endcap of the cavity eliminating PEEK on this side, \emph{iii.} eventually substituting PEEK with sapphire.

To quantify, running such a setup for 1 hour with a quality factor $Q_0 = 3\times 10^5$ at a frequency of 9~GHz and coupling $\beta=2$, would reach an average value of $\approx 2\times 10^{-14}~\text{GeV}^{-1}$, provided that the noise temperature is 0.5~K and the magnet is brought to 9~T. This value is a factor less than 1.5 from the benchmark KSVZ model.

\section*{Acknowledgments}
This work is supported by INFN (QUAX experiment), by the U.S. Department of Energy, Office of Science, National Quantum Information Science Research Centers, Superconducting Quantum Materials and Systems Center (SQMS) under Contract No. DE-AC02-07CH11359, by the European Union’s FET Open SUPERGALAX project, Grant No. 863313, and by the PNRR MUR project number PE0000023-NQSTI. This research was also supported by the Munich Institute for Astro-, Particle and BioPhysics (MIAPbP), which is funded by the Deutsche Forschungsgemeinschaft (DFG, German Research Foundation) under Germany's Excellence Strategy -- EXC-2094 -- 390783311.

This paper is based upon work from COST Action COSMIC WISPers CA21106,
supported by COST (European Cooperation in Science and Technology).

The authors wish to acknowledge the technical support of Anna Calanca, Marco Beatrici, Stefano Lauciani, Maurizio Gatta, Giuseppe Pileggi, Giuseppe Papalino and Daniele Di Bari, as well as of the Mechanical Workshop Department of the Mechanical Design and Construction Service (SPCM) of LNF, and Fabrizio Stivanello and Eduard Chyhyrynets from the National Laboratories of Legnaro for electropolishing the inner surface of the OFHC cavity.


\section{Appendix}
\subsection{Dark Photons}

\begin{figure}
  \centering
      \includegraphics[width=0.97\columnwidth]{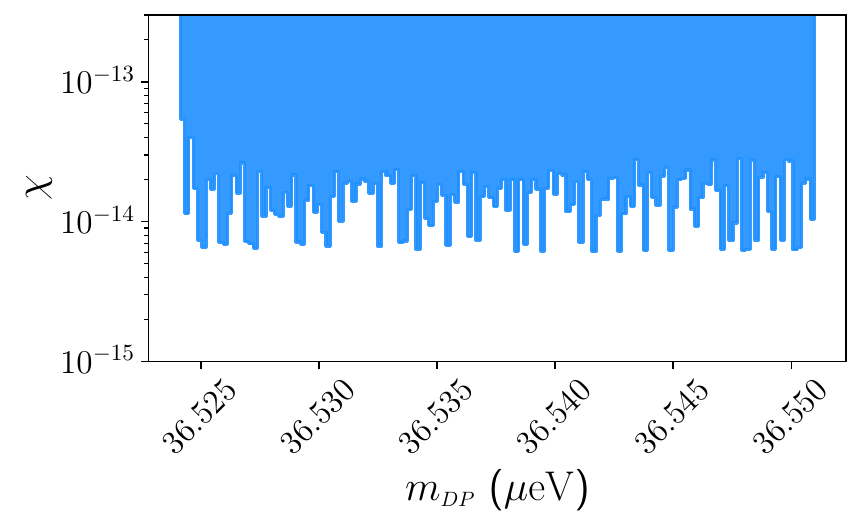}
\caption{\small The 90\% single-sided C.L. upper limit for the kinetic mixing $\chi$ as a function of the DP mass, in the random polarisation scenario.}
\label{fig:DPlimit}
\end{figure}

The Dark Phothon (DP) is weakly coupled to ordinary fields through a small kinetic mixing $\chi$ with the visible photon\cite{caputo2021dark}. We recast the limit on $g_{a\gamma\gamma}$ to one on $\chi$ considering the relation \cite{caputo2021dark,arias2012wispy}:
\begin{equation}
    \chi=g_{a\gamma\gamma}\frac{B}{m_{\scriptscriptstyle \text{DP}}\left|\cos\theta\right|} \, ,
    \label{DP}
\end{equation}
where B is the rms magnetic field value expressed in natural units, $m_{\scriptscriptstyle \text{DP}}$ is the mass of the dark photon, and $\cos\theta=X\cdot B$. The vector $X$ is the polarization of the dark photon field.
Two different scenarios need to be evaluated. In a random polarization scenario $\cos^2\theta$ always assumes the value $1/3$, while in a linear polarization scenario, the value of $\cos\theta$ is an average over the acquisition time. We limit our discussion to the random polarization scenario, since for the second one there is no consensus on which procedure should be used~\cite{caputo2021dark,arias2012wispy}. Fig.~\ref{fig:DPlimit} shows the limits on $\chi$ obtained applying Eq.~\eqref{DP} to the data of Fig.~\ref{fig:gCL}, using $\cos\theta = 1/\sqrt{3}$ .

\subsection{Calibration procedure} \label{sec:calib}
To calibrate the output power, we need to know the gain of the readout line ($L5$). We follow the procedure developed in~\cite{braggio-twpa2022}. The rf diagram of Fig.~\ref{fig:RFdiagram} can  be represented, in dB units, by the following system of three coupled equations:
\begin{eqnarray}
    &S_{51} = G_{L1} + S_{21}^{\scriptscriptstyle \text{cav}}(\nu_c,Q_0,\beta) + G_{L5}, \nonumber\\
    &S_{53} = G_{L3} + S_{22}^{\scriptscriptstyle \text{cav}}(\nu_c,Q_0,\beta) + G_{L5},\label{eq:sistemaEq}\\
    &S_{13} = G_{L1} + S_{12}^{\scriptscriptstyle \text{cav}}(\nu_c,Q_0,\beta) + G_{L3}, \nonumber
\end{eqnarray}
where $S_{51}$, $S_{53}$ and $S_{13}$ are the measured scattering parameters, $G_{L1}$, $G_{L3}$ and $G_{L5}$ are the attenuations/gains of the respective input and output lines, and $S_{ij}^{\scriptscriptstyle \text{cav}}$ are the Lorentzian functions of the cavity (here given in linear form):
\begin{equation}
    S_{21}^{\scriptscriptstyle \text{cav}} = S_{12}^{\scriptscriptstyle \text{cav}} = \left\lvert \frac{2 \sqrt{\beta_1 \beta_2} }{1+\beta_1 +\beta_2 +j Q_0 \delta} \right\rvert,
    \label{eq:S21}
\end{equation}
\begin{equation}
    \begin{split}
    S_{22}^{\scriptscriptstyle \text{cav}} =\,\, &\bigg\lvert \frac{\beta_2^2 - (1+\beta_1)^2 -(Q_0 \delta+q)^2 }{(1+\beta_1+\beta_2)^2 + Q_0^2\delta^2}+\\
    &- j \frac{2\beta_2 Q_0 \delta}{(1+\beta_1+\beta_2)^2 + Q_0^2\delta^2} \bigg\rvert ,
\end{split}
    \label{eq:S22}
\end{equation}
where $\beta_2$ is the coupling to the mobile antenna, $\delta$ is equal to $\omega/\omega_0 - \omega_0/\omega$, and $q$ is a parameter accounting for the small asymmetry in the reflection coefficient.

At each frequency step in Table~\ref{tab:dataset} we perform a simultaneous fit of the $S_{51}$, $S_{53}$ and $S_{13}$ data. Note that in the fits $\beta_1$ is kept fixed at $1.4\times 10^{-3}$, the value expected from simulations, since it was verified to be less than $7\times 10^{-3}$ combining measurements of $L1$ and $L2$ data.

Upon solving Eqs.~\eqref{eq:sistemaEq} using the parameters found from the fit, the gain of the readout line is
\begin{equation}
    G_{L5} = \frac{(S_{51}-S_{21}^{\scriptscriptstyle \text{cav}}) + (S_{53}-S_{22}^{\scriptscriptstyle \text{cav}}) - (S_{13}-S_{12}^{\scriptscriptstyle \text{cav}})}{2},
\end{equation}
evaluated as the average of the spectrum.
The maximum spread of the $G_{L5}$ values within all the subruns is only 0.4~dB, so that we can reasonably consider the gain constant in the evaluated frequency range and equal to the average value of $G_{L5}=70.62$~dB.

\begin{figure}[t]
  \centering
      \includegraphics[width=0.97\columnwidth]{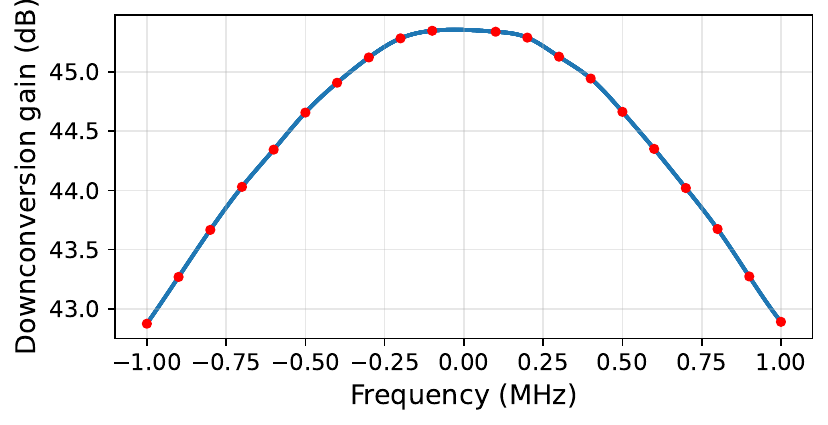}
\caption{\small Calibration of the downconversion electronics in the frequency band of the ADC.}
\label{fig:downconv_gain}
\end{figure}

This value is referred at the input of the splitter. Since the $S$ parameters are measured to and from the VNA ports, to calculate the right value for $G_{L5}$ the contributions of the splitter ($-3$~dB) and of the cable from the splitter to the VNA have been subtracted. Then, a calibration of the downconversion electronics from the splitter input to the DAQ is also necessary. This has been performed only once and is valid for all subruns, since it does not depend on the frequency change. We send a known power at the splitter and measure the downconverted signal at the DAQ input with a Spectrum Analyzer, which gives an absolute power. We repeat the measurement at different LO frequencies to cover all the acquisition bandwidth of the ADC board, from $-1$ to 1~MHz. The calibration curve is shown in Fig.~\ref{fig:downconv_gain}, where the roll-off is due to the ADC internal filters.

Finally, the downconversion calibration curve and the gain $G_{L5}$ are subtracted to the raw power spectra acquired by the ADC to obtain calibrated power spectra (as the one in Fig.~\ref{fig:SG+data}), where the power level is therefore referred at the cavity readout port.

\newpage
\bibliography{biblioFile}


\end{document}